\documentclass[fleqn,twoside]{article}
\usepackage{amsmath}
\usepackage{subfigure}
\usepackage[headings]{espcrc2}

\readRCS
$Id: espcrc2.tex,v 1.2 2004/02/24 11:22:11 spepping Exp $
\usepackage{graphicx}
\usepackage[figuresright]{rotating}

\newcommand{\D}{\mathrm{d}}

\title{
  \quad\vskip-3.3cm \hfill {\normalsize
  \begin{tabular}[t]{l}
                     \rule{0ex}{1ex}MAN/HEP/2009/3 \\[.5ex]
                     \rule{0ex}{1ex}CERN-TH/2009-012
  \end{tabular}}
  \vskip1.5cm 
Soft gluons and superleading logarithms in QCD}

\author{J.R. Forshaw\address[UoM]{School of Physics \&
    Astronomy, University of Manchester, Oxford Road, Manchester M13
    9PL, UK.}\thanks{Talk presented at the workshop ``New Trends in
    HERA Physics'', Ringberg Castle, Tegernsee, 5--10 October 2008.},
  M.H. Seymour\addressmark[UoM]\address{Theoretical Physics Group,
    CERN, CH-1211, Geneva 23, Switzerland.}}

\runtitle{Soft gluons and superleading logarithms in QCD}
\runauthor{J.R Forshaw and M.H. Seymour}

\begin{document}

\begin{abstract}
After a brief introduction to the physics of soft gluons in QCD we present
a surprising prediction. Dijet production in hadron-hadron collisions
provides the paradigm, i.e. $h_1 +h_2 \to jj+X$. In particular, we
look at the case where there is a restriction placed on the
emission of any further jets in the region in between the primary (highest
$p_T$) dijets. Logarithms in the ratio of the jet scale to the veto
scale can be summed to all orders in the strong
coupling. Surprisingly, factorization of collinear emissions fails at
scales above the veto scale and triggers the appearance of double
logarithms in the hard sub-process. The effect appears first at fourth
order relative to the leading order prediction and is subleading in
the number of colours.   
\vspace{1pc}
\end{abstract}

\maketitle

\section{Introducing soft gluons and coherence}
Given a particular short distance process in QCD (``hard scattering'')
we can ask how it will be dressed with additional radiation. A priori,
the question may not be accessible to perturbative QCD because
hadronization effects could wreck the underlying partonic
correlations. However, experiment reveals that the hadronization
process is gentle and we are in business. Our attention should focus
on the most important emissions and these involve either soft gluons
or collinear branchings, or both. By important, we mean that the
suppression provided by the strong coupling is compensated by a large
logarithm, which emerges when the angle between two partons becomes
small or when the energy of a gluon becomes small. 

\begin{figure}[htb]
\centering
\includegraphics[width=15pc]{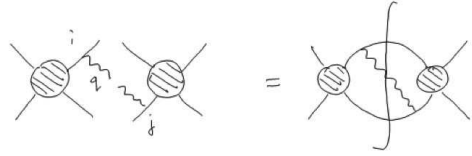}
\caption{Soft gluon emission (wavy line) off a four-parton hard scattering.}
\label{fig:soft1}
\end{figure}

We'll start with a very brief review of soft gluons. The Feynman rule for
the emission of a soft gluon off any fast parton is, to good
approximation, proportional to the four-momentum of the fast
parton. This simplification means that we can factorize the
cross-section for an $n+1$ parton process (i.e. with one soft gluon
and $n$ partons participating in the hard scatter) as
\begin{equation}
\D \sigma_{n+1} = \D \sigma_n \frac{\alpha_s}{2\pi}
\frac{\D E}{E} \frac{\D \Omega}{2\pi} \sum_{ij} C_{ij} E^2
\frac{p_i \cdot p_j}{p_i \cdot q \; p_j \cdot q}
\label{eq:soft}
\end{equation}
where the various symbols are defined (for $n=4$) in
Fig.~\ref{fig:soft1}. The blob in the figure represents a generic
short-distance process and we need only ever consider soft gluon
emissions off external legs since they are incapable of putting the
internal hard propagators on-shell\footnote{Since we use Feynman gauge throughout, we can also neglect self-energy type diagrams, both in the real and virtual contributions, i.e. the sum in Eq.~(1) runs over $i\not=j$.}. The virtual corrections factorize
similarly, and we will have more to say about them shortly. All looks
pretty simple, but there is a major obstacle preventing automated soft
gluon calculations in generic processes: the colour factor $C_{ij}$ is
very difficult to keep track of. Think of emitting many
soft gluons. The leading logarithmic behaviour nests, which means that each emission sees an
effective short-distance process involving all previous
emissions and the corresponding colour structure lives in some large
representation of SU(3). This actually is no problem for theorists who
can calculate in a colour basis independent way \cite{Bonciani:2003nt,Forshaw:2008cq}
but it is a problem when it comes to putting a number on a
cross-section. To date, this problem has not been solved for $n > 3$. Monte Carlo
generators, like HERWIG and PYTHIA, duck the issue by working in the
large $N_c$ approximation. They also exploit the fact that, after
integrating over azimuth in Eq.~(\ref{eq:soft}), soft emissions take place into cones of
successively smaller angles as one moves away from the hard scatter.
The major part of this talk will be concerned with soft gluons but a word first on collinear emissions.

Collinear partonic evolution is easier to deal with. The colour
structure simplifies and partonic evolution reduces to a classical
branching process: it is as if particles are emitted off the parton to
which they are collinear. Hence (after integrating over azimuthal angle)
\begin{equation}
\D \sigma_{n+1} = \D \sigma_n \frac{\alpha_s}{2\pi}
\frac{\D q^2}{q^2} \D z \sum_{ab} P_{ba}(z)
\end{equation}
for $a \to bc$, where parton $b$ carries a fraction $z$ of parton $a$'s momentum, and $P_{ba}(z)$ is the
corresponding splitting function.

The large $N_c$ approximation for
soft emissions permits one to combine them with collinear emissions in
a single parton shower, successive emissions being ordered in
angle. Our task is to go beyond the leading $N_c$ approximation and
gain a better understanding of soft gluon physics.

\begin{figure}[htb]
\centering
\includegraphics[width=15pc]{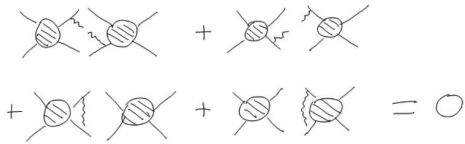}
\caption{Soft gluons cancel in a sum over cuts.}
\label{fig:BN}
\end{figure}

Fig. \ref{fig:BN} is a diagrammatic statement of the Bloch-Nordsieck
Theorem: Summing over cuts, the real and virtual contributions exactly
cancel in the soft gluon approximation. If real emissions are
forbidden for some reason (or contribute to an observable with
anything other than unit weight) then a miscancellation is induced and
that leaves behind a logarithm in the volume $V$ of the phase-space into
which real emission is suppressed. Examples of observables affected in
this way are event shapes (e.g. thrust where $V = 1-T$), production
near threshold ($V = 1 - M^2/s$), Drell-Yan at low $p_T$ ($V =
p_T^2/s$), deep inelastic scattering at large $x$ ($V=1-x$) and ``gaps
between jets''. Our attention now turns to the latter process: it is
the simplest process involving four partons in the hard scatter. Specifically,
one is interested in a final state containing two high $p_T$ jets. The
observable cross-section is defined by imposing that there be no
additional jets in the rapidity region between the two hard jets that have
$p_T > Q_0$. We emphasise that $Q_0$ defines the gap in an
experimentally well defined manner \cite{Adloff:2002em,Chekanov:2006pw}. For this observable $V = Q_0/Q$
where $Q$ is the $p_T$ of the hard jets.

\begin{figure}[htb]
\centering
\includegraphics[width=18pc]{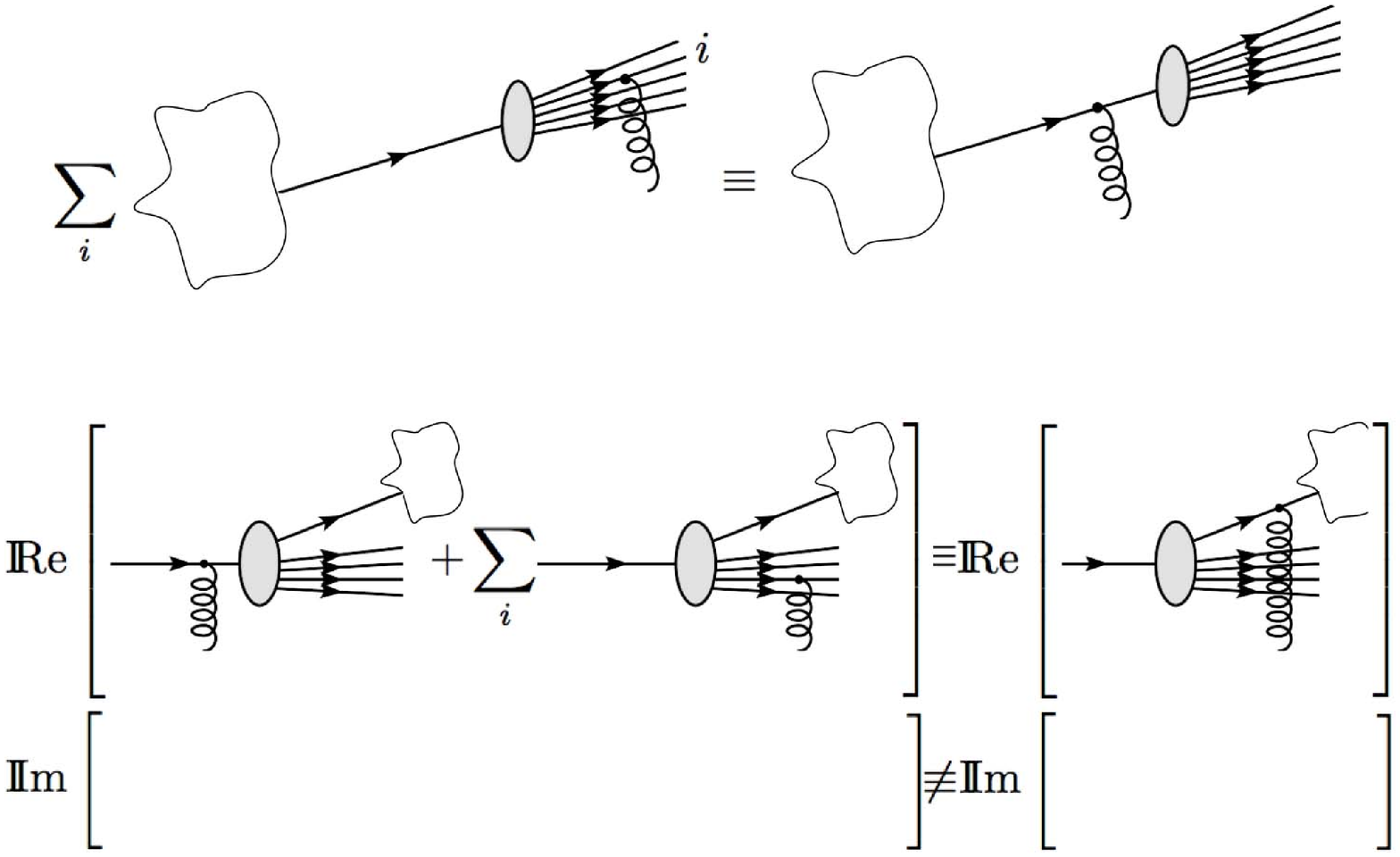}
\caption{Coherence.}
\label{fig:coherence}
\end{figure}

\setcounter{footnote}{0}
QCD coherence underpins the angular ordering of soft gluon emissions
and can be used to argue that a wide-angle soft gluon emitted off a
bunch of collinear partons can always be re-attached to the primary
hard parton that defines the direction of the bunch. It is a statement
at the amplitude level for the virtual soft gluon corrections\footnote{and the
  cross-section level for real emissions.} and is illustrated in
Fig.~\ref{fig:coherence}. It is understood that there are any
number of additional partons exiting the hard-scatter blob (not
shown) and that the soft gluon re-attaches to one of them. The
crucial point to note is that coherence defined in this diagrammatic
sense breaks down for the imaginary part of the amplitude. It is
spoilt by virtual soft gluons that put on-shell the partons to which
they attach. Such gluons are variously called Coulomb or Glauber
gluons in the literature and they are the bane of those wanting to
prove QCD factorization theorems. Assuming coherence leads us to conclude that
the gaps-between-jets observable should contain only single
logarithms, i.e. terms $\sim \alpha_s^n \ln^n(Q/Q_0)$, since it is
inclusive over the collinear regions.

\begin{figure}[htb]
\centering
\includegraphics[width=3.5pc]{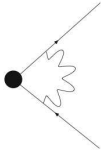}
\caption{The soft gluon dressing of a two-parton process.}
\label{fig:epem}
\end{figure}

\section{Pre 2001: Exponentiation}

We start with the simplest example: $e^+ e^- \to q \bar{q}$ with a restriction $p_T
< Q_0$ on the emission of additional jets into a region of rapidity
$Y$ somewhere in between the primary jets. Naively, we need only compute the diagram shown
in Fig.~\ref{fig:epem}, where the blob represent the short distance
physics. The soft gluon should be integrated over the region into
which real emissions are forbidden. In particular, the real part of
this one-loop amplitude should be integrated over rapidities in the
gap and $k_T > Q_0$ since the complementary region is cancelled by
real emissions according to Bloch-Nordsieck. We refer to the soft
gluons that generate this real part as eikonal gluons. The imaginary part is
more subtle: it arises as a result of Coulomb gluon exchange and has
nothing to cancel against (it makes no sense to speak of the rapidity
of these gluons). To pick up all of the leading logarithms to all orders in
$\alpha_s$ one might think that we need only iterate the process of adding a
soft gluon between the jets, with each successive emission occurring at
much smaller $k_T$ than the one before\footnote{We shall see in the
  next section that this does not sum all of the leading logarithms.}. It is as if all prior
emissions are sitting in the short-distance blob in
Fig.~\ref{fig:epem}. The net effect is an amplitude
\begin{eqnarray}
A &=& A_0 \exp\left( -\frac{\alpha_s}{\pi} \int_{Q_0}^{Q} \frac{\D
    k_t}{k_t} C_F (Y+\rho_\mathrm{jet}) \right) \nonumber \\ &\times& \exp\left(+\frac{2\alpha_s}{\pi} \int_{0}^{Q} \frac{\D
    k_t}{k_t} C_F \; i \pi \right)~,
\end{eqnarray}
where $\rho_\mathrm{jet}$ is some function dependent upon the jet
algorithm (we have ignored the running of the coupling but it is easy enough
to restore it). The colour structure in this two-jet example is simple (the $q\bar{q}$
must be in a colour singlet) and as a result the Coulomb gluon ($i
\pi$) term generates an unimportant phase.

\begin{figure}[htb]
\centering
\subfigure[]{\includegraphics[width=0.2\textwidth]{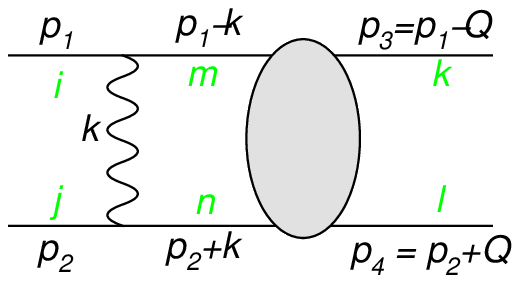}}
\subfigure[]{\includegraphics[width=0.2\textwidth]{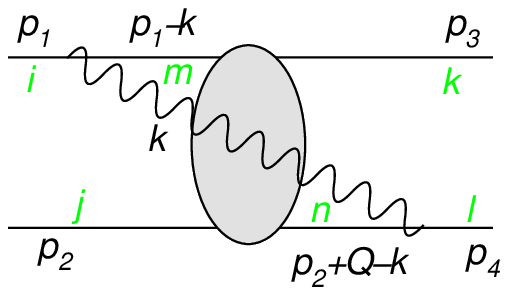}} \\
\subfigure[]{\includegraphics[width=0.2\textwidth]{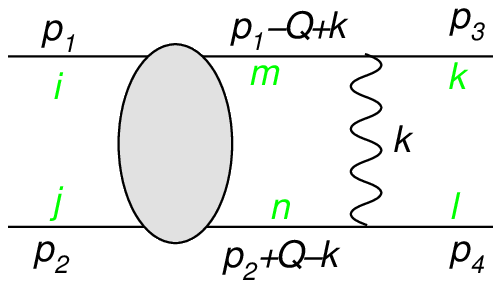}}
\subfigure[]{\includegraphics[width=0.2\textwidth]{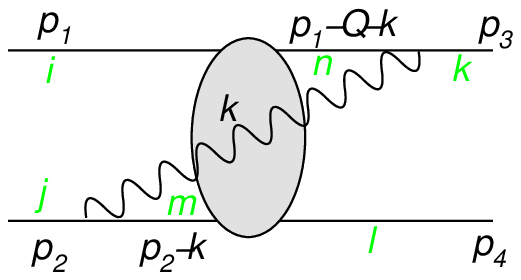}} 
\caption{The soft gluon dressing of a four-parton process.}
\label{fig:hh}
\end{figure}

The story is much the same for three and four-parton processes. In
hadron-hadron collisions we need to compute the virtual corrections
illustrated in Fig.~\ref{fig:hh}. These are the leading diagrams at
large $Y$ and there are two others (where the soft gluon links partons 1
and 3 or partons 2 and 4), which contribute only to the dependence upon $\rho_{\mathrm{jet}}$.
The only subtlety arises because the colour structure is now not so
simple. For quark-quark scattering one can think of projecting the
amplitude onto a specific basis and the soft gluons can generate
mixing. The result is an amplitude that goes as
\begin{equation}
\mathbf{A} = \exp\left( -\frac{2\alpha_s}{\pi} \int_{Q_0}^{Q} \frac{\D
    k_t}{k_t} \boldsymbol{\Gamma} \right) \mathbf{A}_0~.
\end{equation}
In the $t$-channel singlet-octet basis (e.g. the top left entry generates pure singlet evolution):
\begin{equation}
  \mathbf{\Gamma} =
  \left(\begin{array}{cc}
    \frac{N^2-1}{4N}\rho_\mathrm{jet} & \frac{\sqrt{N^2-1}}{2N}i\pi \\
    \frac{\sqrt{N^2-1}}{2N}i\pi & -\frac1Ni\pi + \frac N2Y
    +\frac{N^2-1}{4N}\rho_\mathrm{jet}\end{array}\right) \label{eq:gamma4}
\end{equation}
and $\sigma = \mathbf{A}^\dag \mathbf{A}$. Now the Coulomb gluons
don't only lead to a phase: witness the $i \pi$ terms in the
evolution matrix. The inclusive nature of the region $k_t < Q_0$ does
however mean that they cancel in that region since the evolution is
determined by a purely imaginary matrix, i.e. $\Im
\mathrm{m} \; \boldsymbol{\Gamma}$. Hence the universal lower cut-off
of $Q_0$. The calculation of these Sudakov logarithms in all $2 \to 2$
sub-processes can be found in \cite{Oderda:1998en,Berger:2001ns}.

Although we have been focussing our attention on gaps-between-jets, it
is worth recalling that the formalism carries over almost without change to
the case when a colour singlet particle(s) is produced in between the
jets (e.g. a Higgs boson) since the soft gluon evolution is blind to
it. The only change occurs in the jet function because the final state
jets now recoil against the Higgs \cite{Forshaw:2007vb}. Notice that any
studies of a jet veto in Higgs-plus-two-jets that are performed using the
general purpose Monte Carlos necessarily miss the effects of Coulomb
gluons and hence any singlet-octet mixing. This is a particularly
serious deficiency at large $Y$ and for stringent vetos, where colour singlet
exchange will eventually dominate.    

\section{2001-2006: Non-global logarithms}

\begin{figure}[htb]
\centering
\includegraphics[width=15pc]{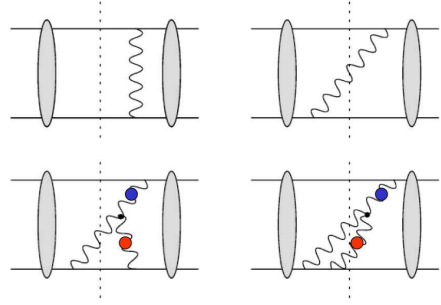}
\caption{The origin of non-global logarithms.}
\label{fig:ngl}
\end{figure}

The exponentiation of soft gluons is not the whole story. In
2001, Dasgupta \& Salam \cite{Dasgupta:2001sh,Dasgupta:2002bw}
realized that a whole tower of leading
logarithms was being neglected. Fig.~\ref{fig:ngl} illustrates the problem.
The top two panes illustrate the real-virtual cancellation of
Bloch-Nordsieck and the bottom two illustrate the problem. The blue (darker)
blob indicates a soft gluon that is emitted outside of the
gap but with $p_T > Q_0$. In the left picture, it
receives a virtual correction from the red gluon, which necessarily has
a lower $p_T$ but is still above $Q_0$. Crucially, this red gluon can
be in the gap region. The right picture represents the real emission
that would cancel the aforementioned virtual correction in a
sufficiently inclusive observable. However, the real emission diagram
is forbidden by the gap definition (since the red gluon is ``in the
gap and has $p_T > Q_0$"), which means that the virtual correction on
the left has nothing to cancel against. Clearly the argument applies
to any number of real emissions outside of the gap and instead of
being able to focus only on virtual corrections to the primary hard
scatter we now have to contemplate amplitudes with arbitrary numbers
of real emissions and the virtual corrections to them. That is clearly
a tall order and, to date, these leading ``non-global'' logarithms
have only been summed up to all orders within the leading $N_c$
approximation and in the case of two-parton
observables.\footnote{Estimates have been made for $2 \to 2$ processes
  in \cite{Appleby:2003sj}.}
Physical insight into this effect can be gained by noting that the
all-orders result is essentially that, by requiring the gap region to be
free of radiation above $Q_0$, a region outside it must also be
depopulated, named the buffer zone in \cite{Dasgupta:2002bw}.
A very interesting feature worth mentioning is that the equation that
sums these logarithms \cite{Banfi:2002hw} is identical in form to the Balitsky-Kovchegov
equation \cite{Balitsky:1995ub,Kovchegov:1999yj} in small-$x$ physics
\cite{Marchesini:2003nh,Weigert:2003mm}. Is this
coincidence or does it reflect a deeper connection between high
energy scattering amplitudes and jet physics? The physics behind non-global logarithms
is almost embarrasingly simple: the observable is obviously sensitive
to the fact that emissions outside of the gap cannot emit back into
the gap.

\section{Post 2006: Super-leading logarithms?}
As a first step to understanding non-global logarithms we
decided to compute the tower of leading non-global logarithms that
arise as a result of one gluon sitting outside the gap \cite{Forshaw:2006fk}. One might view
this as the first\footnote{Or second, if we count the global (exponentiating)
  series as the contribution from ``zero gluons outside of the gap''.} term in an expansion in the number of out-of-gap
gluons and it has the virtue that we can go ahead and calculate the
cross-section without making the large $N_c$ approximation. To make
progress we need two new ingredients: (i) we need to know how to emit a
real (soft) gluon off the four-parton amplitude; (ii) we need to be able
to dress the resulting five-parton amplitude with a virtual soft gluon.
\begin{figure}[htb]
\centering
\includegraphics[width=10pc]{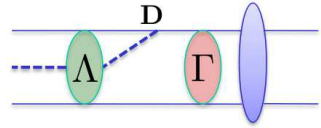}
\caption{Five-parton evolution.}
\label{fig:five}
\end{figure}
Fig.~\ref{fig:five} illustrates schematically how the five-parton
matrix element is computed. Assuming that successive emissions occur
at progressively lower $k_T$, starting from the hard scatter at scale
$Q$ (on the right in the figure), the general amplitude contains first a period of virtual
evolution, which is carried out by successive operations of the
four-parton evolution matrix $\boldsymbol{\Gamma}$, followed by the
emission of a real gluon (denoted by $\mathbf{D}$ in the figure) and this in
turn is followed by a period of five-parton evolution (denoted by
$\boldsymbol{\Lambda}$). Kyrieleis \& Seymour computed
$\mathbf{D}$ and $\boldsymbol{\Lambda}$ for four-quark amplitudes
\cite{Kyrieleis:2005dt} and Sj\"odahl recently computed the remaining five parton
evolution matrices in a specific colour basis \cite{Sjodahl:2008fz}. It is also
possible to write down the evolution matrix for a general $n$-parton
amplitude in a basis-independent manner  \cite{Bonciani:2003nt,Forshaw:2008cq}. The result
is (with a particular choice of the overall phase):
\begin{eqnarray}
  \mathbf{\Gamma} &=&
  \frac12Y\mathbf{T}_t^2
  +i\pi\mathbf{T}_1\cdot\mathbf{T}_2
  +\frac14\sum_{i\in F}\rho_{\mathrm{jet},i} \; \mathbf{T}_i^2
\nonumber\\&+&
  \frac12\sum_{(i<j)\in L}\lambda_{\mathrm{jet},ij} \;
  \mathbf{T}_i\cdot\mathbf{T}_j \nonumber \\ &+&
  \frac12\sum_{(i<j)\in R}\lambda_{\mathrm{jet},ij} \;
  \mathbf{T}_i\cdot\mathbf{T}_j ~.\label{eq:gamma}
\end{eqnarray}
$\mathbf{T}_i$ is the colour charge of parton $i$ (partons 1 and 2 are
incoming and the others are outgoing) and $\mathbf{T}_t$
is the net charge exchanged in the $t$-channel, i.e.  $\mathbf{T}_t =
\sum_{i \in L} \mathbf{T}_i = - \sum_{i \in R} \mathbf{T}_i$ and the
sums are over all incoming and outgoing partons lying on either the
left or right side of the gap.
Notice that the terms involving $\rho_{\mathrm{jet},i}$ are Abelian since
$\mathbf{T}_i^2 = C_F \boldsymbol{1}$ or $C_A \boldsymbol{1}$
depending upon whether parton $i$ is a quark/antiquark or
gluon. $\lambda_{\mathrm{jet},ij}$ is a second jet function.

Now $\mathbf{\Gamma}$ has a very important property: it
is safe against final state collinear singularities.  More specifically, if any
two or more partons in the final state become collinear with each
other, the soft gluon evolution of the system is identical to the evolution 
of the system in which the collinear partons are replaced by a single parton with the
same total colour charge.  That is, if $k$ and $l$ are the collinear partons, then
$\mathbf{\Gamma}$ depends only upon $\mathbf{T}_k+\mathbf{T}_l$ and not upon the
$\mathbf{T}_k$ or $\mathbf{T}_l$ separately. A proof can be found
in \cite{Forshaw:2008cq} and results in the factorization expressed in the upper
diagram of Fig.~\ref{fig:coherence}. This is to be contrasted with the
initial state collinear limit, i.e. in which one or more outgoing
partons becomes collinear with one of the incoming partons. The proof
clearly breaks down for the imaginary part because of the Coulomb
gluon term $i \pi \mathbf{T}_1 \cdot \mathbf{T}_2$, which depends only
on the colours of the initial state partons and not on the sum of the
colour charges of the collinear partons. As a result the factorization is broken, as illustrated in
the lower part of Fig.~\ref{fig:coherence}. It is this breakdown of naive
coherence that leads directly to the appearance of super-leading
logarithms in the calculation of the gaps-between-jets cross-section.

For one gluon outside of the gap the cross-section is thus
\begin{equation}
  \sigma_1 = -\frac{2\alpha_s}{\pi}\int_{Q_0}^Q\frac{dk_T}{k_T}
  \int_{\mathrm{out}}\frac{dy\,d\phi}{2\pi}
  \Bigl(\Omega_R+\Omega_V\Bigr), \label{eq:outside}
\end{equation}
where the integral is over the phase-space for the out-of-gap gluon and
\begin{eqnarray}
  \label{masterV}
  \Omega_V &=& \mathbf{A}_0^\dagger
  \exp\left(-\frac{2\alpha_s}{\pi}\int_{Q_0}^Q\frac{dk_T'}{k_T'}
    \mathbf{\Gamma}^\dagger\right)
\nonumber\\&&
  \exp\left(-\frac{2\alpha_s}{\pi}\int_{Q_0}^{k_T}\frac{dk_T'}{k_T'}
    \mathbf{\Gamma}\right)
  \boldsymbol{\gamma} \nonumber \\ &&
  \exp\left(-\frac{2\alpha_s}{\pi}\int_{k_T}^Q\frac{dk_T'}{k_T'}
    \mathbf{\Gamma}\right)
  \mathbf{A}_0+\mathrm{c.c.},\\
  \label{masterR}
  \Omega_R &=& \mathbf{A}_0^\dagger
  \exp\left(-\frac{2\alpha_s}{\pi}\int_{k_T}^Q\frac{dk_T'}{k_T'}
    \mathbf{\Gamma}^\dagger\right)
  \mathbf{D}_\mu^\dagger \nonumber \\ &&
  \exp\left(-\frac{2\alpha_s}{\pi}\int_{Q_0}^{k_T}\frac{dk_T'}{k_T'}
    \mathbf{\Lambda}^\dagger\right)
\nonumber\\&&
  \exp\left(-\frac{2\alpha_s}{\pi}\int_{Q_0}^{k_T}\frac{dk_T'}{k_T'}
    \mathbf{\Lambda}\right)
  \mathbf{D}^\mu \nonumber \\ &&
  \exp\left(-\frac{2\alpha_s}{\pi}\int_{k_T}^Q\frac{dk_T'}{k_T'}
    \mathbf{\Gamma}\right)
  \mathbf{A}_0.
\end{eqnarray}
For $\Omega_R$, this has the structure advertised in
Fig.~\ref{fig:five}. $\Omega_V$ involves virtual corrections
to the four-parton matrix element and $\boldsymbol{\gamma}$ adds
the virtual out-of-gap gluon; after integrating over rapidity and
azimuth it is equal to $\Re \mathrm{e} \; \mathbf{\Gamma}$ using
Eq.~(\ref{eq:gamma4}). 

Rather than numerically evaluate $\sigma_1$, we shall compute it in
the particular case that the out-of-gap gluon is collinear to one of
the incoming partons. Naively, the integral over the rapidity of the
out-of-gap gluon yields a divergence (it can have arbitrarily large
rapidity). Traditionally that would not pose any problem because
colour coherence (and the ``plus prescription'') would say that
$\Omega_R +\Omega_V=0$ in this limit, however in light of our previous
discussion we are interested to see how the ``coherence breaking''
Coulomb gluon contribution works out. We need to treat the collinear
region with a little more care than we have done. In particular we
must go beyond the soft gluon approximation in order to account for
the fact that a gluon of $k_T > Q_0$ cannot have infinite rapidity and
finite energy. We should instead treat the collinear region using the
collinear (but not soft) approximation. The integral over rapidity
ought then to be replaced by
\begin{eqnarray}
\int\limits_{\text{out}}dy~~\left.  \frac{d\sigma}{dyd^{2}%
k_{T}}\right\vert _{\text{soft}}&\rightarrow&
 \nonumber \\  & & \hspace*{-4cm}
\int\limits^{y_{\text{max}}}dy~\left.  \frac{d\sigma}{dyd^{2}k_{T}}\right\vert
_{\text{soft}}+\int\limits_{y_{\text{max}}}^{\infty}dy~\left.  \frac{d\sigma
}{dyd^{2}k_{T}}\right\vert _{\text{collinear}}  .\nonumber \\  \label{eq:softcol}%
\end{eqnarray}
In this equation $y_{\text{max}}$ divides the regions in
which the soft and collinear approximations are used and the dependence on it
will cancel in the sum. Now we know that
\begin{eqnarray}
\int\limits_{y_{\text{max}}}^{\infty}dy~\left.  \frac{d\sigma}{dyd^{2}k_{T}
}\right\vert _{\text{collinear}}&=& \nonumber \\ && \hspace*{-4.4cm} \int\limits_{y_{\text{max}}}^{\infty
}dy~\left(  \left.  \frac{d\sigma_{\text{R}}}{dyd^{2}k_{T}}\right\vert
_{\text{collinear}}+\left.  \frac{d\sigma_{\text{V}}}{dyd^{2}k_{T}}\right\vert
_{\text{collinear}}\right)~, \nonumber \\ &&
\end{eqnarray}
where the contribution due to real gluon emission is%
\begin{eqnarray}
\int\limits_{y_{\text{max}}}^{\infty}dy~\left.  \frac{d\sigma_{\text{R}}
}{dyd^{2}k_{T}}\right\vert _{\text{collinear}}~  &=& \nonumber \\ &&
\hspace*{-4cm} \int\limits_{0}
^{1-\delta}dz\frac{1}{2}\left(  \frac{1+z^{2}}{1-z}\right)  \frac
{q(x/z,\mu^{2})}{q(x,\mu^{2})}A_{\text{R}}\nonumber\\
&& \hspace*{-4.5cm} =\int\limits_{0}^{1-\delta}dz\frac{1}{2}\left(  \frac{1+z^{2}}{1-z}\right)
\left(  \frac{q(x/z,\mu^{2})}{q(x,\mu^{2})}-1\right)  A_{\text{R}}
\nonumber \\ && \hspace*{-3cm}
+\int\limits_{0}^{1-\delta}dz\frac{1}{2}\frac{1+z^{2}}{1-z}A_{\text{R}}
\label{eq:col}
\end{eqnarray}
and the contribution due to virtual gluon emission is
\begin{eqnarray}
\int\limits_{y_{\text{max}}}^{\infty}dy~\left.  \frac{d\sigma_{\text{V}}
}{dyd^{2}k_{T}}\right\vert _{\text{collinear}} &=& \nonumber \\ && \hspace*{-2cm} 
\int\limits_{0}^{1-\delta
}dz\frac{1}{2}\left(  \frac{1+z^{2}}{1-z}\right)
A_{\text{V}}~.
\end{eqnarray}
In Eq.~(\ref{eq:col}), $q(x,\mu^{2})$ is the parton distribution function for a
quark in a hadron at scale $\mu^{2}$ and momentum fraction $x$. The factors
$A_{\text{R}}$ and $A_{\text{V}}$ contain the $z$ independent factors which
describe the soft gluon evolution and the upper limit on the $z$ integral is
fixed since we require $y>y_{\text{max}}$\footnote{The approximation arises
since we assume for simplicity that $\Delta y$ is large and $\delta$ is small.
This approximation does not affect the leading behaviour and can easily be
made exact if necessary.}:
\begin{equation}
\delta\approx\frac{k_{T}}{Q}\exp\left(  y_{\text{max}}-\frac{\Delta y}%
{2}\right)  .
\end{equation}
We are prepared for the fact that $A_{\text{R}}+A_{\text{V}}\neq0$
(and we shall see it explicitly in Eq.~(\ref{eq:nonzero})) but if it were the case that
$A_{\text{R}}+A_{\text{V}}=0$ then the virtual emission contribution would
cancel identically with the corresponding term in the real emission
contribution leaving behind a term regularised by the `plus prescription'
(since we can safely take $\delta\rightarrow0$ in the first term of
Eq.~(\ref{eq:col})). This term could then be absorbed into the evolution of the
incoming quark parton distribution function by choosing the factorisation
scale to equal the jet scale $Q$.

The miscancellation therefore induces an additional contribution of the form
\begin{eqnarray}
\int\limits_{0}^{1-\delta}dz\frac{1}{2}\left(  \frac{1+z^{2}}{1-z}\right)
(A_{\text{R}}+A_{\text{V}})  &=&  \\ && \hspace*{-4cm}
\ln\left(  \frac{1}{\delta}\right)
(A_{\text{R}}+A_{\text{V}})+\text{subleading}\nonumber \\ && \hspace*{-5cm}
\approx\left(  -y_{\text{max}}+\frac{\Delta y}{2}+\ln\left(  \frac{Q}
{k_{T}}\right)  \right)  (A_{\text{R}}+A_{\text{V}}). \nonumber
\end{eqnarray}
The $y_{\text{max}}$ dependence cancels with that coming from the soft
contribution in Eq.~(\ref{eq:softcol}) leaving only the logarithm. The
leading effect of treating properly the collinear region is therefore
simply to introduce an effective upper limit to the integration over
rapidity in Eq.~(\ref{eq:outside}), i.e.
\begin{eqnarray}
\int_{Q_0}^Q \frac{\D k_T}{k_T} \int_{\mathrm{out}} \D y & \to &
\nonumber \\ &&
\hspace*{-3cm}
\int_{Q_0}^Q \frac{\D k_T}{k_T} \Bigg( \int^{y_\mathrm{max}} \D y +
(-y_\mathrm{max}+\ln \frac{Q}{k_T}) \Bigg) \nonumber \\ &&
\hspace*{0cm} = \frac{1}{2} \ln^2 \frac{Q}{Q_0}.
\end{eqnarray}
In this way the super-leading logarithm emerges. We see that the
non-collinear region is sub-leading and hence use the fact that,
in the collinear limit, the
evolution matrices simplify so that (neglecting an overall Abelian factor) the cross-section
reduces to \cite{Forshaw:2008cq}
\begin{eqnarray}
  \label{fullresult}
  \sigma_1 &=& -\frac{2\alpha_s}{\pi}\int_{Q_0}^Q\frac{dk_T}{k_T}
  \left(2\ln\frac{Q}{k_T}\right) \nonumber \\ &&
  \Biggl\langle\!m_0\Biggr|
  e^{-\frac{2\alpha_s}{\pi}\int_{k_T}^Q\frac{dk_T'}{k_T'}
    \left(\frac12Y\mathbf{t}_t^2
    -i\pi\mathbf{t}_1\cdot\mathbf{t}_2\right)}
\nonumber\\ &&
  \Biggl\{
  \mathbf{t}_1^2
  e^{-\frac{2\alpha_s}{\pi}\int_{Q_0}^{k_T}\frac{dk_T'}{k_T'}
    \left(\frac12Y\mathbf{t}_t^2
    -i\pi\mathbf{t}_1\cdot\mathbf{t}_2\right)} \nonumber \\ &&
  e^{-\frac{2\alpha_s}{\pi}\int_{Q_0}^{k_T}\frac{dk_T'}{k_T'}
    \left(\frac12Y\mathbf{t}_t^2
    +i\pi\mathbf{t}_1\cdot\mathbf{t}_2\right)}
\nonumber\\&&
  -\mathbf{t}_1^{a \dagger}
  e^{-\frac{2\alpha_s}{\pi}\int_{Q_0}^{k_T}\frac{dk_T'}{k_T'}
    \left(\frac12Y\mathbf{T}_t^2
  -i\pi\mathbf{T}_1\cdot\mathbf{T}_2\right)} \nonumber \\ &&
  e^{-\frac{2\alpha_s}{\pi}\int_{Q_0}^{k_T}\frac{dk_T'}{k_T'}
    \left(\frac12Y\mathbf{T}_t^2
  +i\pi\mathbf{T}_1\cdot\mathbf{T}_2\right)}
  \mathbf{t}_1^a
  \Biggr\}
\nonumber\\&&
  e^{-\frac{2\alpha_s}{\pi}\int_{k_T}^Q\frac{dk_T'}{k_T'}
    \left(\frac12Y\mathbf{t}_t^2
    +i\pi\mathbf{t}_1\cdot\mathbf{t}_2\right)}
  \Biggl|m_0\!\Biggr\rangle.
\end{eqnarray}
We have changed to a bra-ket notation where
$|m_0\rangle$ denotes the lowest order matrix element: it better suits
the basis independent language wherein we are to think of the colour
charges as objects that map vectors in $m$-parton colour space to
vectors in $(m+1)$-parton colour space. Consequently, we have used
lower ($\mathbf{t}_i$) and upper ($\mathbf{T}_i$) case symbols to
represent the charge of parton $i$ in the four and five parton
systems respectively. Note that it is the non-commutativity of $\mathbf{T}_t^2$ and
$\mathbf{T}_1\cdot\mathbf{T}_2$ (and similarly $\mathbf{t}_t^2$ and
$\mathbf{t}_1\cdot\mathbf{t}_2$) that prevents this expression from
cancelling to zero: if they commuted then the two exponentials could be
combined, all $\mathbf{T}_1\cdot\mathbf{T}_2$ and
$\mathbf{t}_1\cdot\mathbf{t}_2$ dependence would cancel, $\mathbf{t}_1$
could be commuted through $\mathbf{T}_t^2$ and the real and virtual
parts would be identical. For fewer than four external partons colour
conservation means that we can
always write $\mathbf{T}_1 \cdot \mathbf{T}_2 \propto \boldsymbol{1}$
and hence the Coulomb gluons only generate an unimportant phase.

We can get a better idea of what is going on if we expand the $\left\{
    ... \right\}$ in Eq.~(\ref{fullresult}) order by order in $\alpha_s$. By setting
the exponentials to unity outside this bracket we are assuming that
the out-of-gap gluon has the largest $k_T$ (we shall look at the other
possibilities shortly). The first two orders vanish identically:
\begin{equation}
  \Biggl\{\phantom{...}\Biggr\}_0=
  \mathbf{t}_1^2-\mathbf{t}_1^{a \dagger}\mathbf{t}_1^a = 0.
\end{equation}
\begin{equation}
  \Biggl\{\phantom{...}\Biggr\}_1 =
  -\frac{2\alpha_s}{\pi}\int_{Q_0}^{k_T}\frac{dk_T'}{k_T'}
  \Biggl\{\mathbf{t}_1^2Y\mathbf{t}_t^2
  -\mathbf{t}_1^{a \dagger} Y\mathbf{T}_t^2\,\mathbf{t}_1^a\Biggr\}
\end{equation}
is also zero because 
\begin{equation}
  \mathbf{T}_t^2\,\mathbf{t}_1^a=\mathbf{t}_1^a\,\mathbf{t}_t^2.
\end{equation}
Expanding to order $\alpha_s^2$ yields
\begin{eqnarray}
  \Biggl\{\phantom{...}\Biggr\}_2 &=&
  \left( \frac{i\pi Y}{2}\right)
  \left(-\frac{2\alpha_s}{\pi}\int_{Q_0}^{k_T}\frac{dk_T'}{k_T'}\right)^2
\nonumber \\ && \hspace*{-1cm}
  \Biggl\{
  \mathbf{t}_1^2
    \left[\mathbf{t}_t^2\, , \mathbf{t}_1\cdot\mathbf{t}_2 \right]
  -\mathbf{t}_1^{a\dagger}
    \left[\mathbf{T}_t^2\,, \mathbf{T}_1\cdot\mathbf{T}_2 \right]
  \mathbf{t}_1^a \label{eq:a2}
  \Biggr\} .  \nonumber \\ &&
\end{eqnarray}
Note that this result comes only from the case where there is one
Coulomb gluon and one eikonal gluon  either side of the cut. Now, this
term is not zero but the corresponding matrix element is zero,
i.e. $\langle m_0 | \left\{ ~ \right\}_2 | m_0 \rangle = 0.$

At the next order we obtain a non-zero result:
\begin{eqnarray}
\!
  \Biggl\{\phantom{...}\Biggr\}_3 &\equiv&
  -\frac{Y\pi^2}{6}
  \left(-\frac{2\alpha_s}{\pi}\int_{Q_0}^{k_T}\frac{dk_T'}{k_T'}\right)^3
  \nonumber \\ &&
  \Biggl\{
  \mathbf{t}_1^2
   \Bigl[ \left[
    \mathbf{t}_t^2\,,\mathbf{t}_1\cdot\mathbf{t}_2\right] ,
      \mathbf{t}_1\cdot\mathbf{t}_2 \Bigr]  \nonumber\\ && \hspace*{-1cm}
  -\mathbf{t}_1^{a \dagger}
  \Bigl[ \left[
    \mathbf{T}_t^2\,, \mathbf{T}_1\cdot\mathbf{T}_2\, \right]  ,
      \mathbf{T}_1\cdot\mathbf{T}_2 \Bigr]
     \mathbf{t}_1^a
  \Biggr\}. \label{eq:nonzero}
\end{eqnarray}
Substituting back into Eq.~(\ref{fullresult}), evaluating the colour
matrix element and performing the transverse momentum integrals, we obtain a
contribution to the first super-leading logarithm from configurations in
which the out-of-gap gluon is hardest of
\begin{eqnarray}
&& \hspace*{-0.7cm}\sigma_{1,\mathrm{hardest}}=-\sigma_0 \left(
  \frac{2\alpha_s}{\pi}\right)^4 \ln^5 \left(\frac{Q}{Q_0}\right)
\pi^2 Y \frac{N^2 -2}{240} \nonumber \\ && 
\end{eqnarray}
in the case of $qq \to qq$. The other subprocesses lead to
different colour matrix elements and hence a different coefficient of
the form $(a N^2 + b)$. The explicit results are presented in \cite{Forshaw:2008cq}.
\begin{figure}[t]
\centering
\subfigure[]{\includegraphics[width=0.2\textwidth]{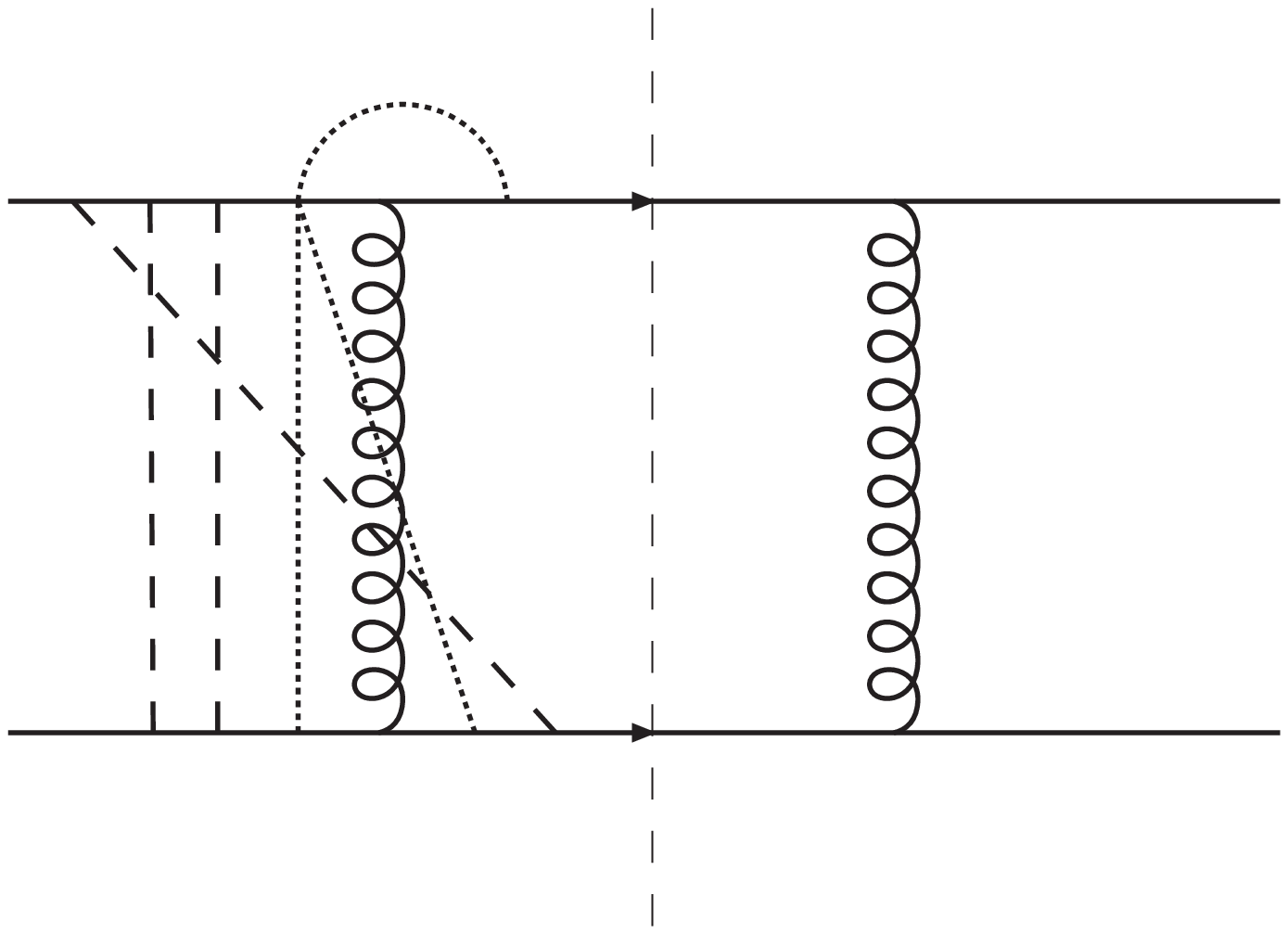}}
\subfigure[]{\includegraphics[width=0.2\textwidth]{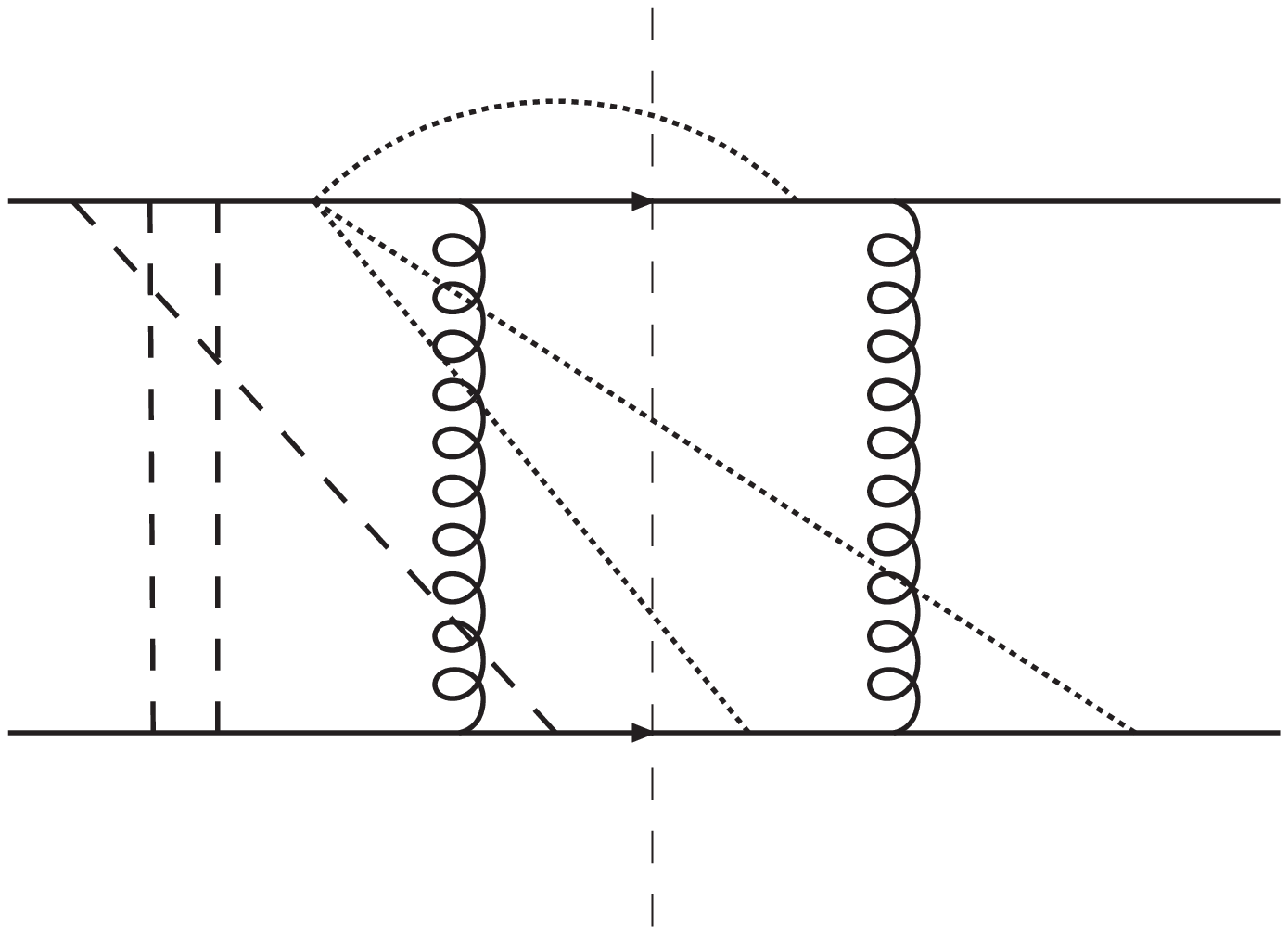}} \\
\subfigure[]{\includegraphics[width=0.2\textwidth]{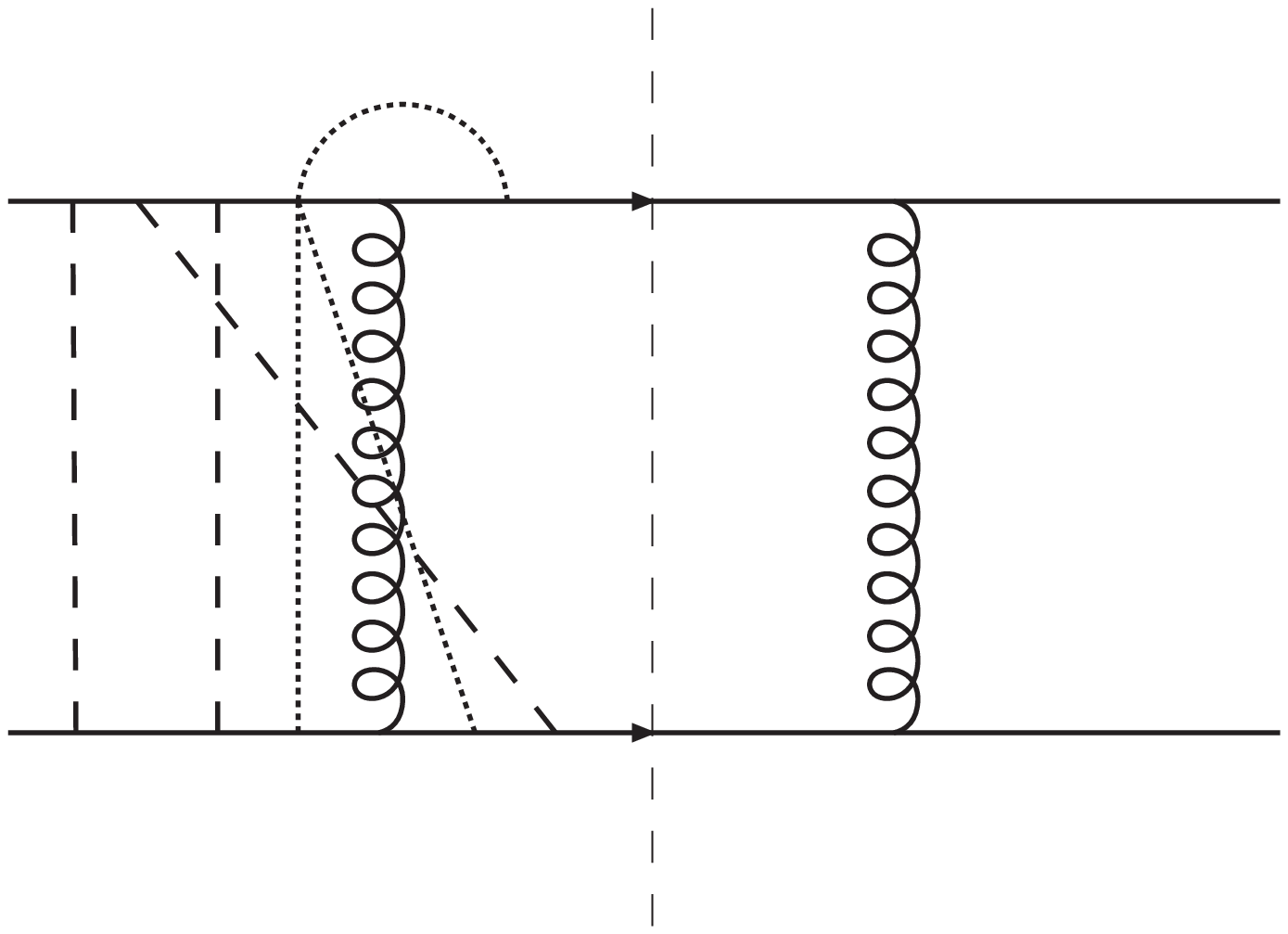}}
\subfigure[]{\includegraphics[width=0.2\textwidth]{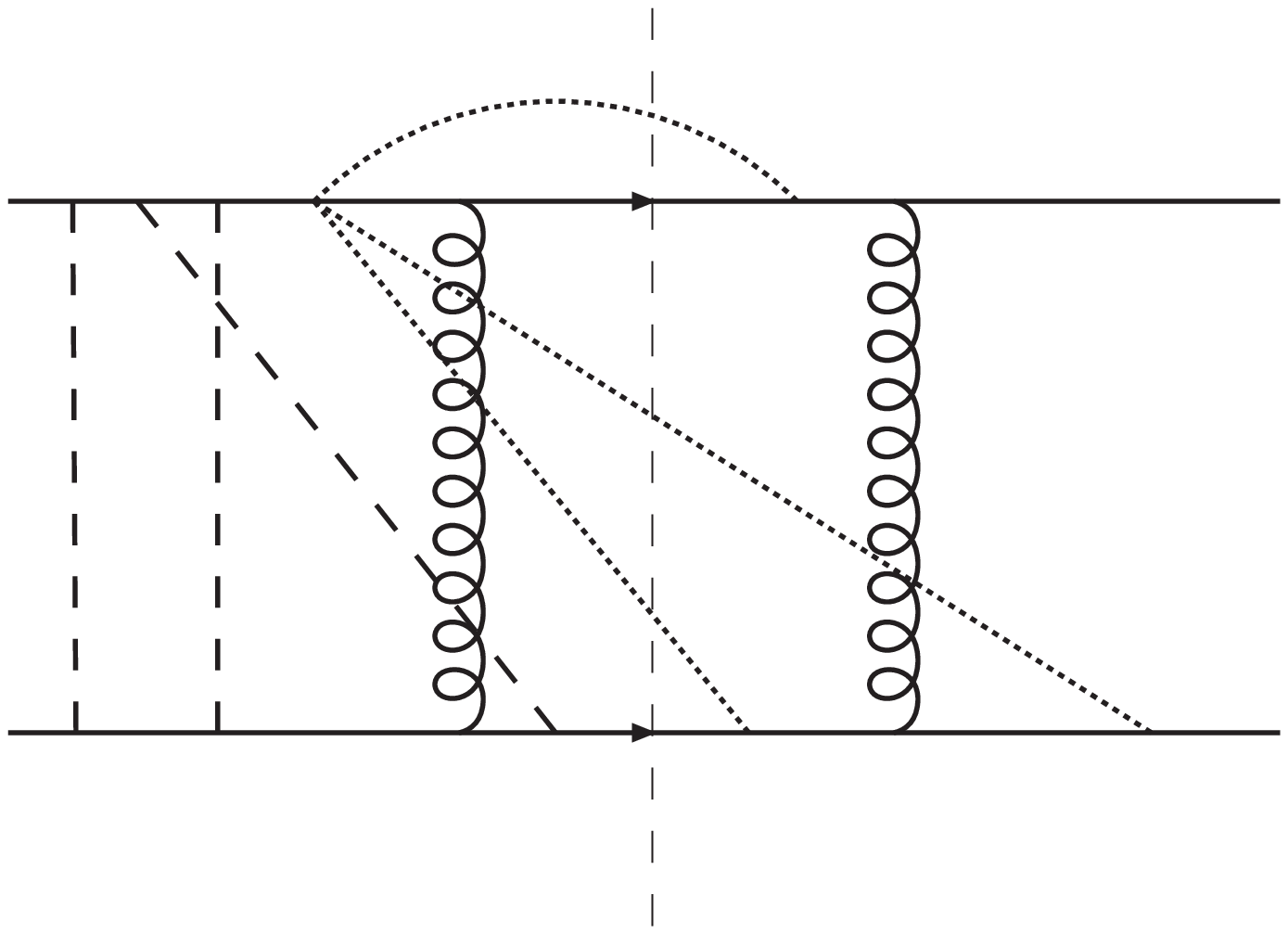}} \\
\subfigure[]{\includegraphics[width=0.2\textwidth]{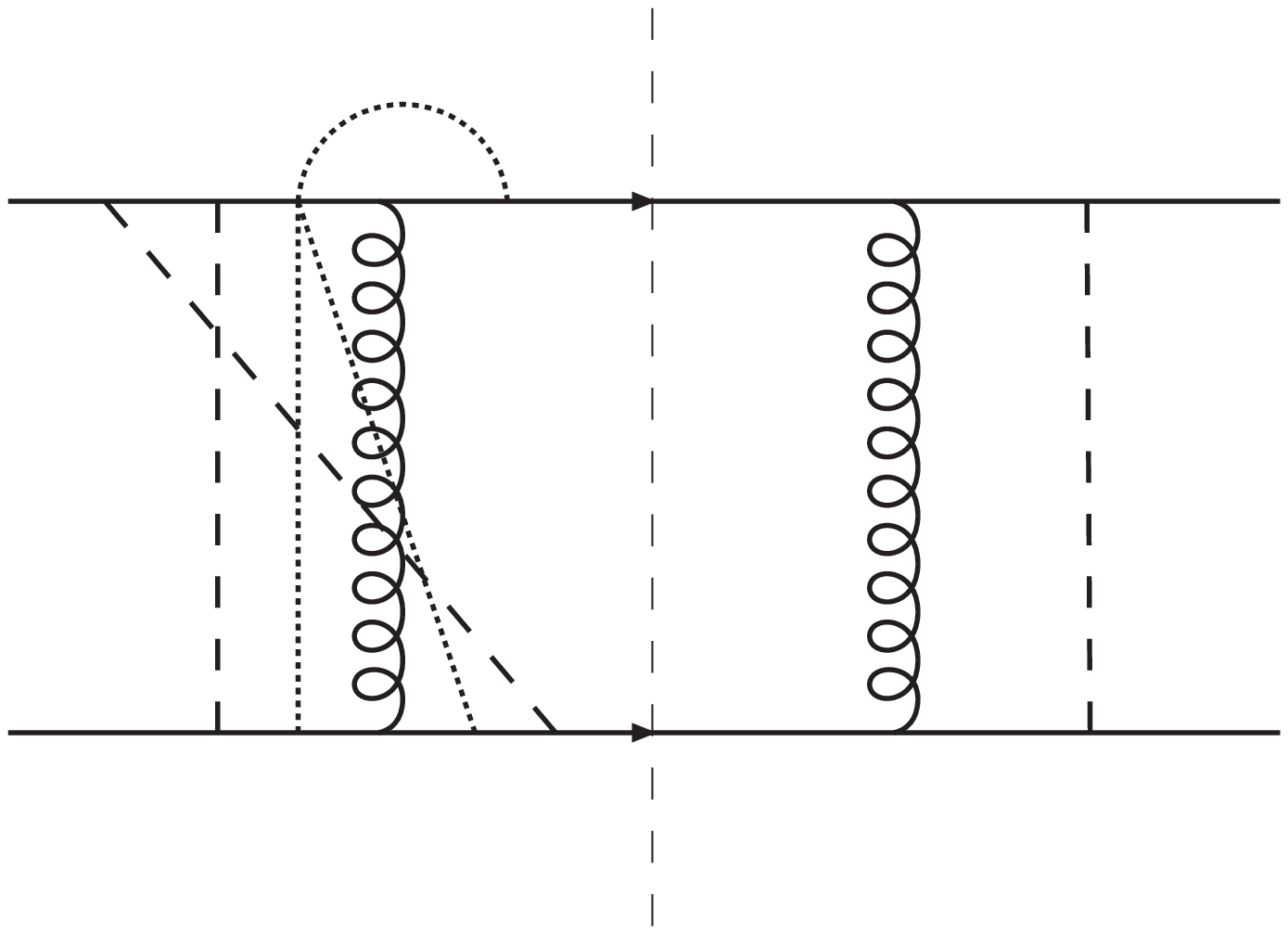}}
\subfigure[]{\includegraphics[width=0.2\textwidth]{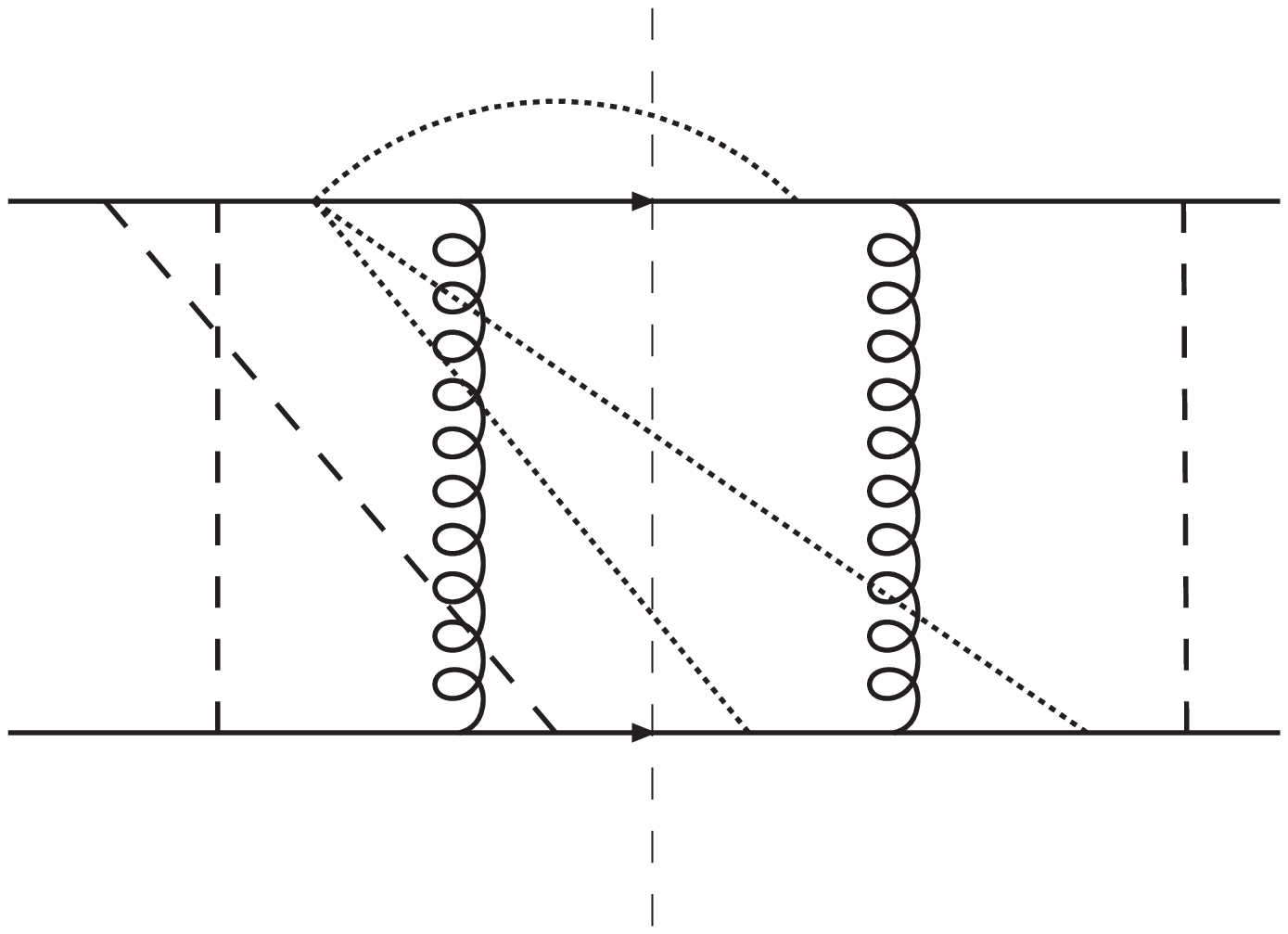}} \\
\subfigure[]{\includegraphics[width=0.2\textwidth]{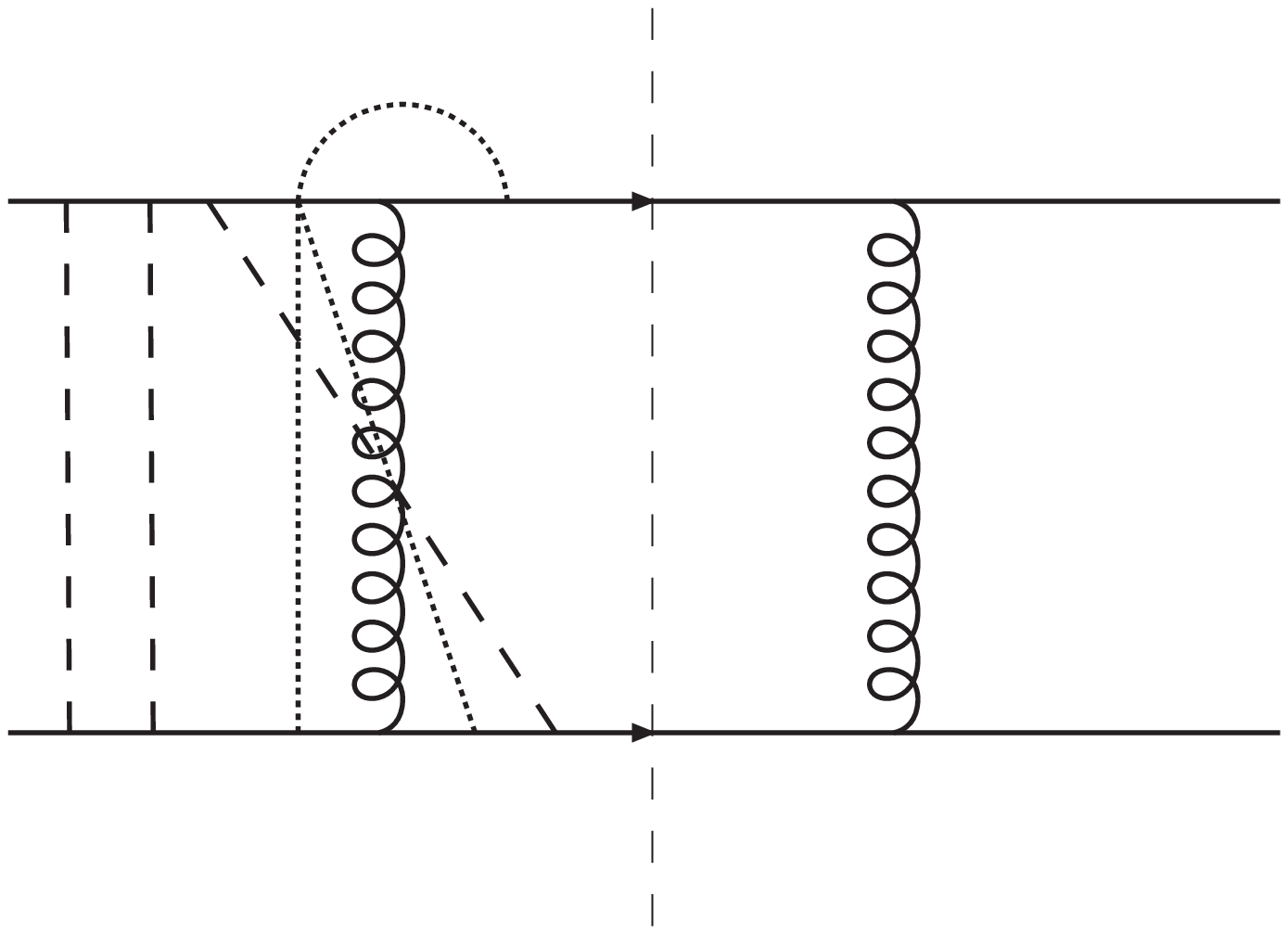}}
\subfigure[]{\includegraphics[width=0.2\textwidth]{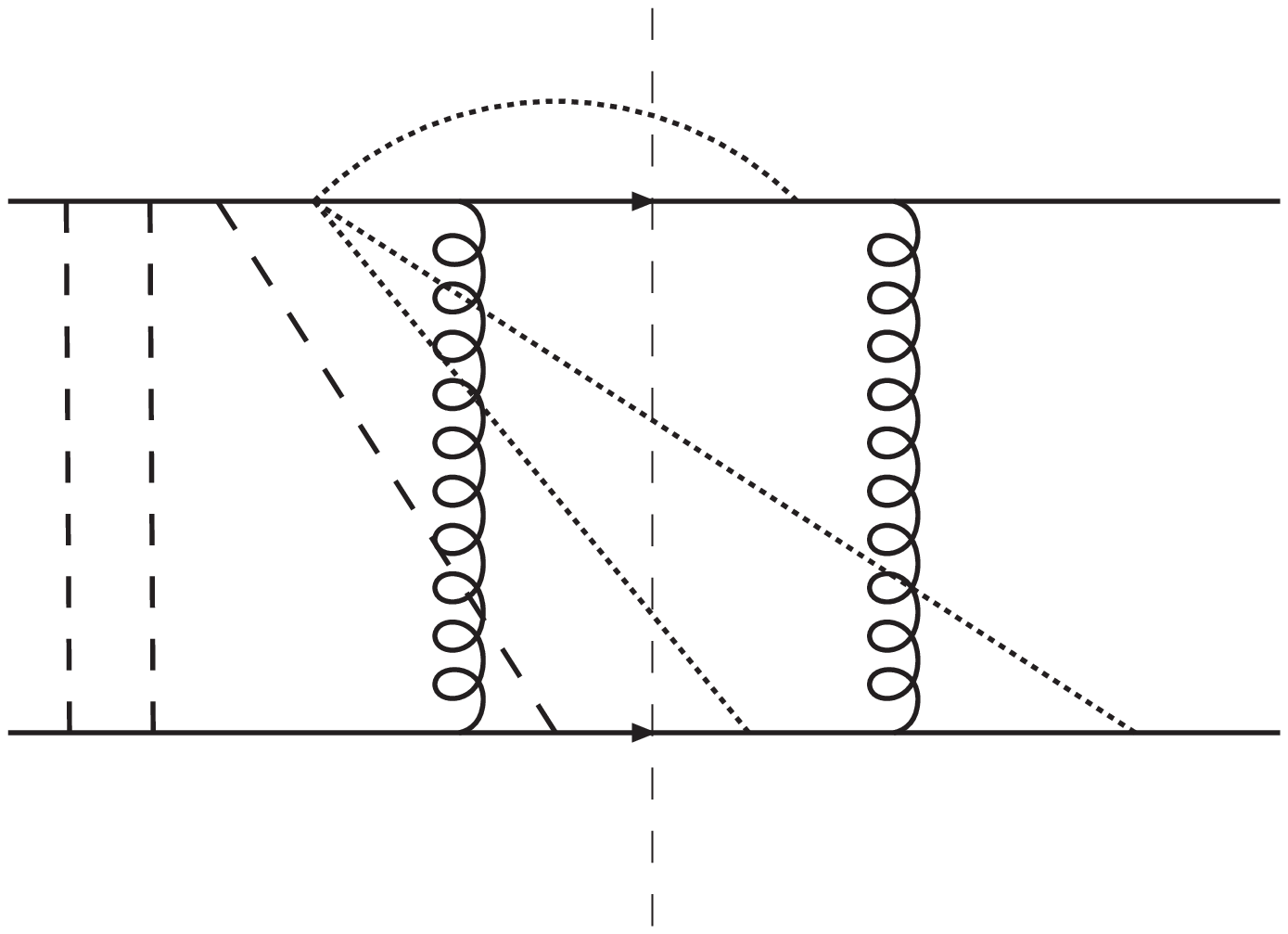}} \\
\caption{The relevant Feynman diagrams in the case that the out-of-gap (dotted) gluon is the hardest gluon. The dashed lines indicate soft (eikonal and Coulomb) gluons. Each subfigure represents three Feynman diagrams, corresponding to the three different ways of attaching the
out-of-gap gluon. In diagrams~(e) and~(f) the soft gluon to the right of the
cut should only be integrated over the region in which it has transverse
momentum less than the out-of-gap gluon.}
\label{fig:hardest}
\end{figure}

To complete the calculation of the order $\alpha_s^4$ super-leading
logarithm we need to compute the contribution arising from the case
where there is just one virtual emission of higher $k_T$ than the out-of-gap emission. 
Now we use the expression for  $\{~\}_2$ derived in Eq.~(\ref{eq:a2}) in conjunction
with the order $\alpha_s$ expansion of the exponential factors that
lie outside of the main bracket in Eq.~(\ref{fullresult}). Note that
this is the only remaining contribution to the lowest order
super-leading logarithm since all lower order expansions of the main
bracket in Eq.~(\ref{fullresult}) (i.e. $\{ ~ \}_1$ and $\{ ~ \}_0$)
vanish. The result is
\begin{eqnarray}
\sigma_{1,\mathrm{second-hardest}}&=& \nonumber \\ && \hspace*{-2.5cm}
-\sigma_0 \left(
  \frac{2\alpha_s}{\pi}\right)^4 \ln^5 \left(\frac{Q}{Q_0}\right)
\pi^2 Y \frac{N^2 -1}{120}. \nonumber \\ && 
\end{eqnarray}
At order $\alpha_s^4$ relative to the lowest order result, the total
super-leading contribution to the gaps-between-jets cross-section for
$qq \to qq$ arising from one soft gluon emission outside of the gap is therefore 
\begin{eqnarray}
&& \hspace*{-0.7cm}\sigma_{1}=-\sigma_0 \left(
  \frac{2\alpha_s}{\pi}\right)^4 \ln^5 \left(\frac{Q}{Q_0}\right)
\pi^2 Y \frac{3N^2 -4}{240}. \nonumber \\ && 
\end{eqnarray}
 
\begin{figure}[t]
\centering
\subfigure[]{\includegraphics[width=0.2\textwidth]{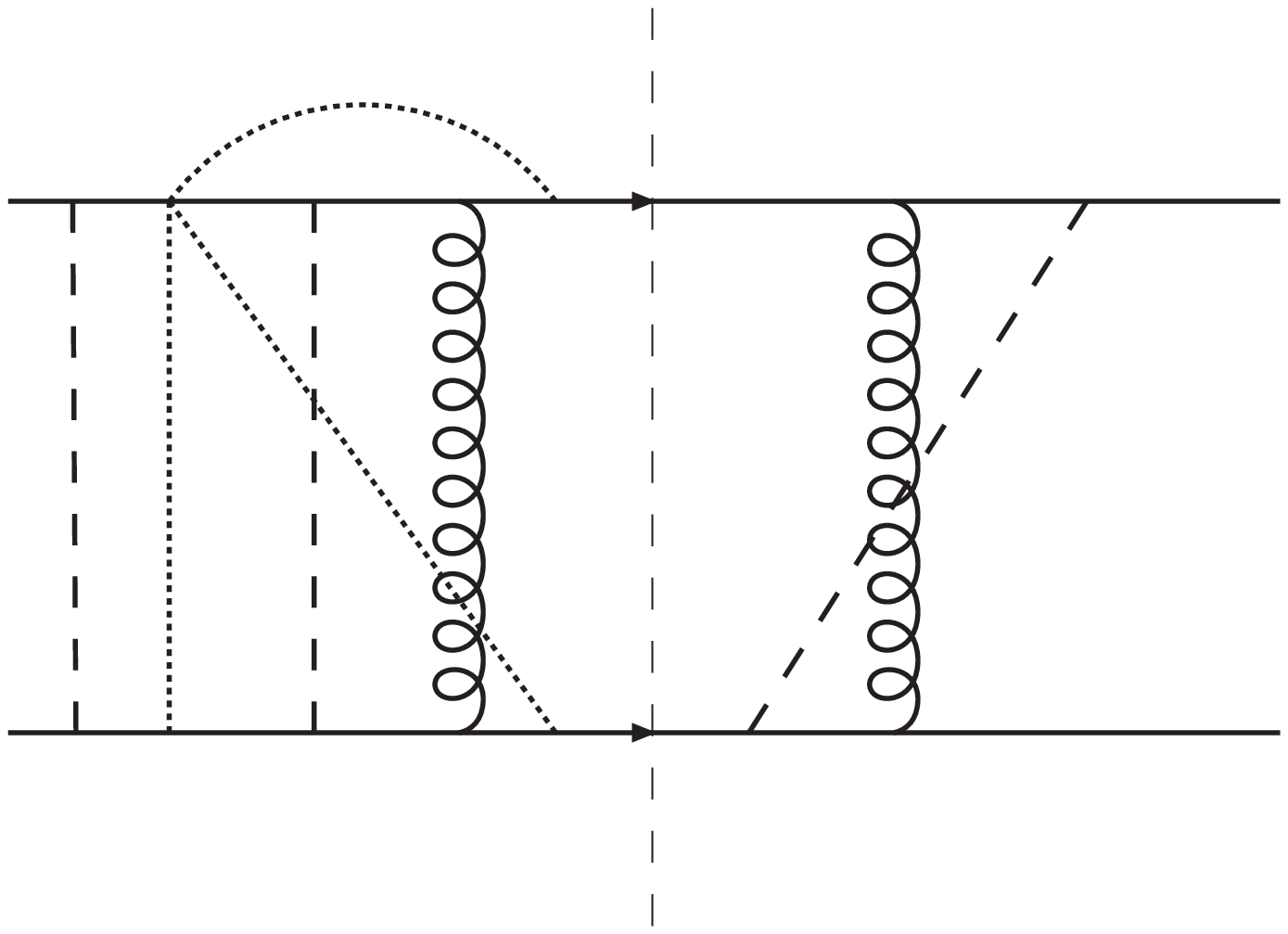}}
\subfigure[]{\includegraphics[width=0.2\textwidth]{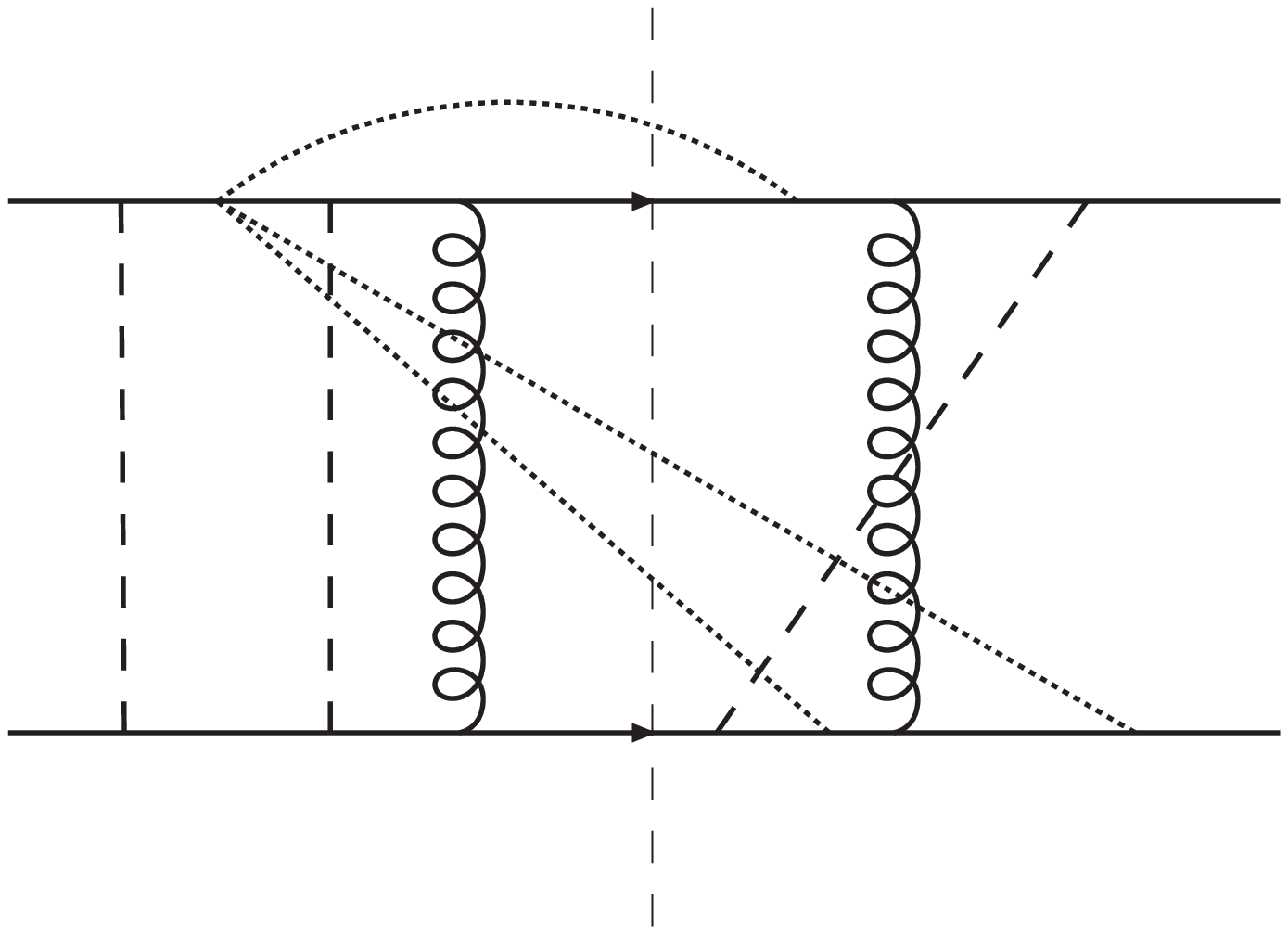}} \\
\subfigure[]{\includegraphics[width=0.2\textwidth]{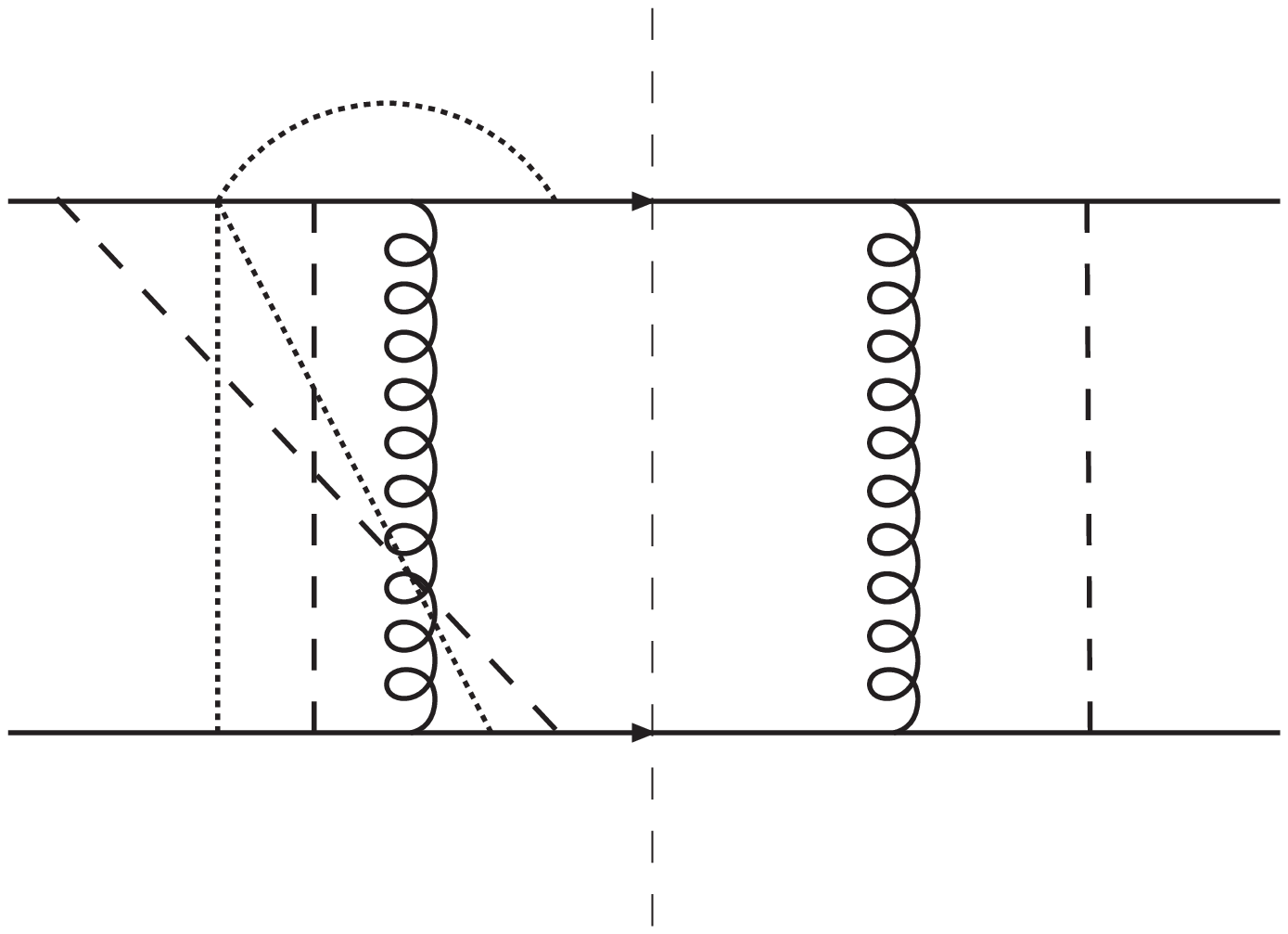}}
\subfigure[]{\includegraphics[width=0.2\textwidth]{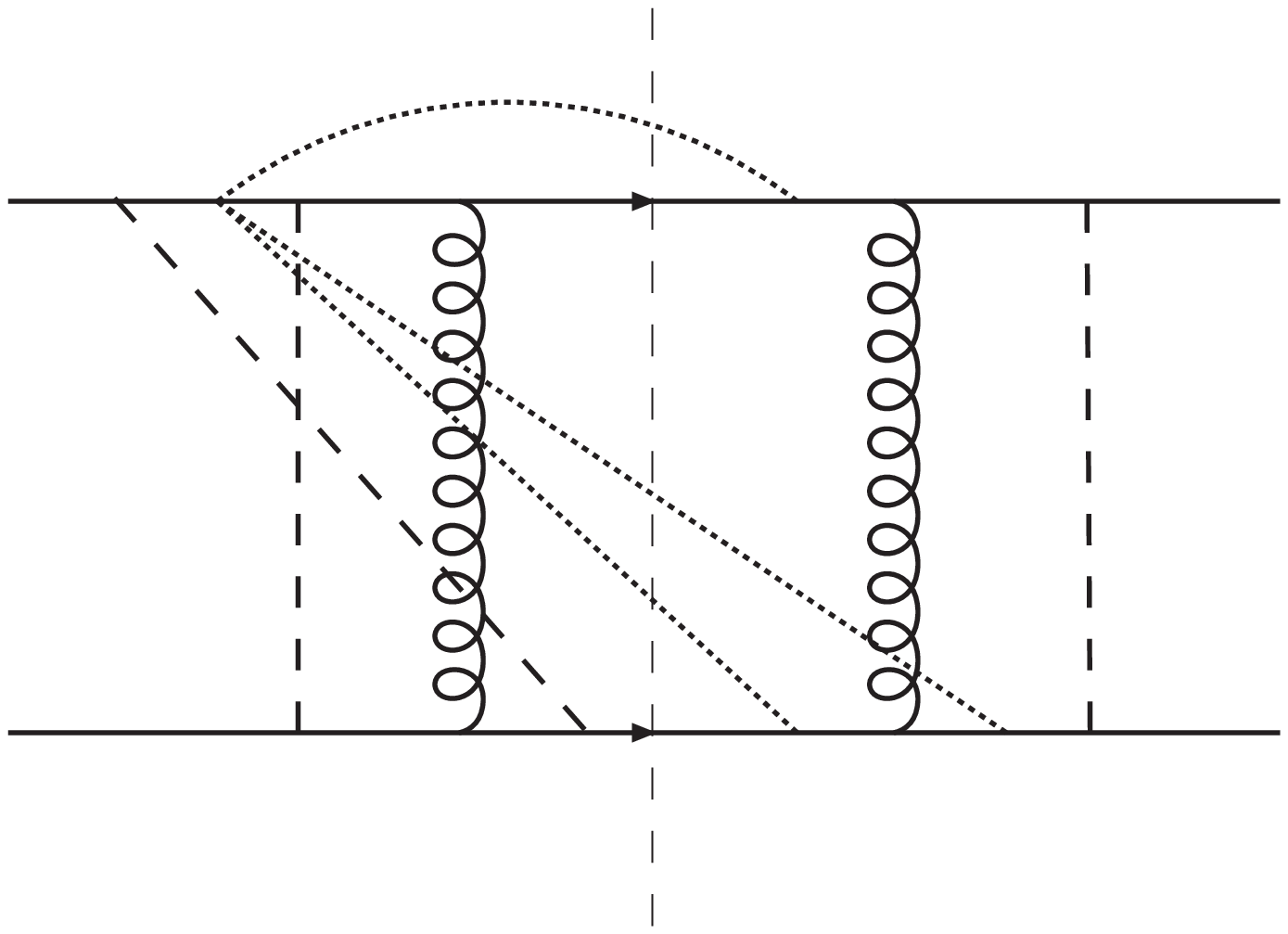}} 
\caption{The relevant Feynman diagrams in the case that the out-of-gap (dotted) gluon is the second hardest gluon. The dashed lines indicate soft (eikonal and Coulomb) gluons. Each subfigure represents three Feynman diagrams, corresponding to the three different ways of attaching the out-of-gap gluon.
The soft gluon to the right of the cut should only be integrated over the
region in which it has transverse momentum less than the out-of-gap
gluon.}
\label{fig:nexttohardest}
\end{figure}
Of the other sub-processes, the result for $qg \to qg$ is worth singling out
because the size of the super-leading log does not depend upon whether
the out-of-gap gluon is collinear to the quark or to the gluon.

\section{Outlook}
\label{outlook}

\begin{figure}[t]
\centering
\includegraphics[width=0.4\textwidth]{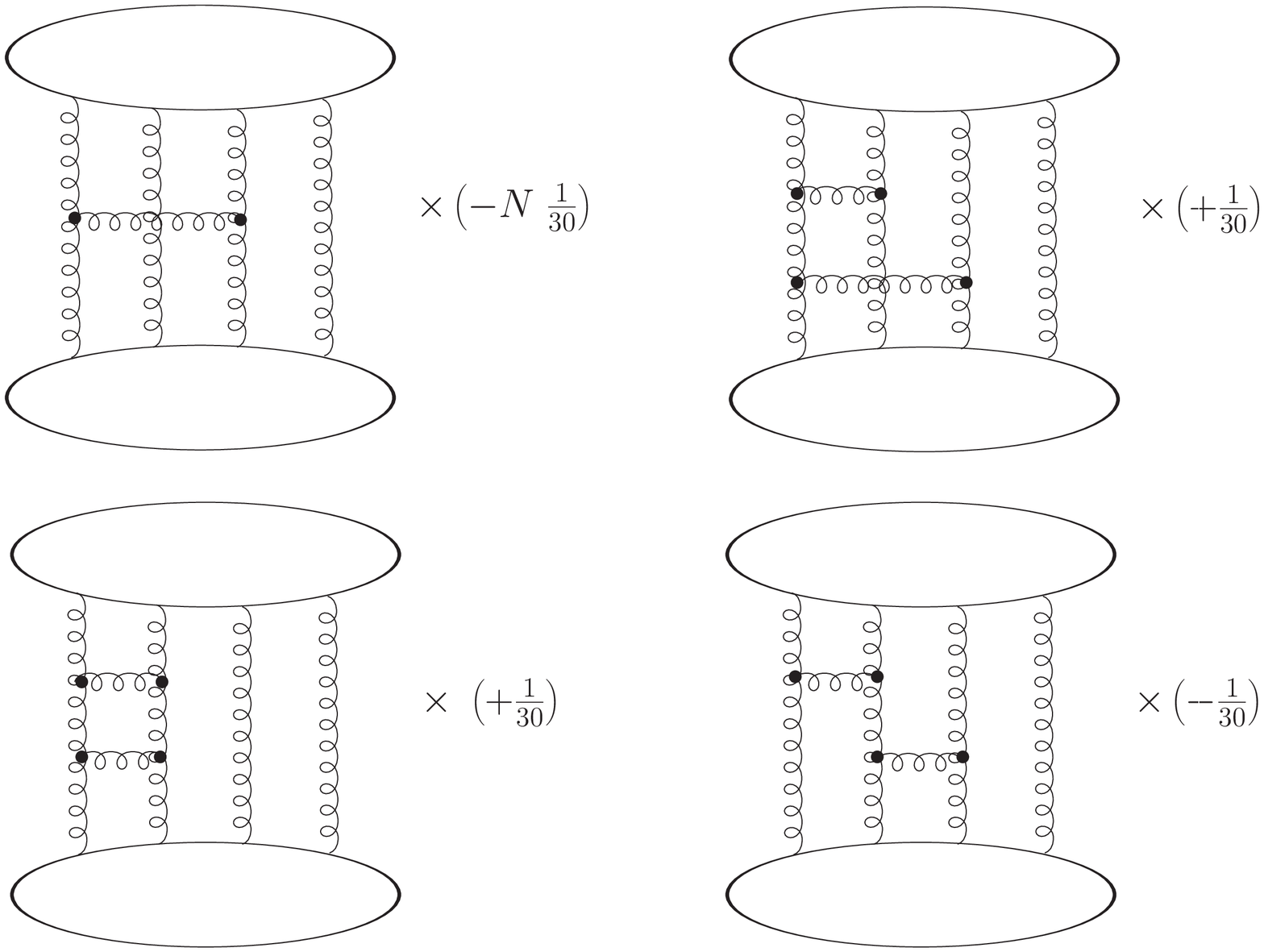}
\caption{The four diagrams that generate the colour matrix elements
  when the out-of-gap gluon is the hardest gluon. In the case that the
out-of-gap gluon is next-to-hardest, only the first diagram
contributes. The upper and lower loops can be quarks, anti-quarks or
gluons. Note that these are not Feynman diagrams or even uncut
diagrams; they represent only the colour factor of the final result.
In the first diagram one of the original six gluon lines has been contracted away, resulting in the additional factor of $N_c$.}
\label{fig:traces}
\end{figure}

The super-leading logarithms constitute a surprising breakdown of collinear
factorization in an observable that sums inclusively over the
collinear regions: the `plus prescription' fails to operate and double
(soft-collinear) logarithms make their appearance. The effect is due
to Coulomb (Glauber) gluon exchange and it arises only in observables
with at least four coloured partons in the hard scatter, is
non-Abelian, sub-leading in $N_c$ and appears first at fourth order in
the strong coupling relative to the lowest order. The
implications for the gaps-between-jets cross-section are clear: collinear
logarithms can be summed into the parton density functions only up to scale
$Q_{0}$ and the logarithms in $Q/Q_{0}$ from further collinear evolution must
be handled seperately. Moreover, since we now have a source of double
logarithms, the calculation of the single logarithmic series necessarily
requires knowledge of the two-loop evolution matrices \cite{Aybat:2006wq,MertAybat:2006mz}.

Some questions still remain open however: the structure
of higher order super-leading logarithms; how widespread they are in
other observables for hadron collisions; and whether they can be
reorganized and resummed or removed by a suitable redefinition of
observables or of incoming partonic states. 

We close by noting that further simplifications are possible. In particular, the
commutators between gluon exchanges from an external leg in different
orders can be written as emission off the exchanged Coulomb gluons. The
colour matrix elements can then be shown to be identical to those
illustrated in 
Fig.~\ref{fig:traces}. Once again it is clear that the super-leading
logarithms arise as a result of the non-Abelian nature of the Coulomb gluon
interaction. We also see that the coefficient of the
superleading logarithm is independent of whether the out-of-gap gluon
is collinear to parton~1 or~2, because the result is invariant under
interchange of the particle types in the upper and lower loops. Colour
structures like these are reminiscent of small-$x$ physics.

\section*{Acknowledgements}

We thank Stefano Catani, Mrinal Dasgupta, James Keates, Simone Marzani, Malin
Sj\"odahl and George Sterman for interesting discussions of these and
related topics. Thanks also to the workshop organizers, both for
inviting JRF to deliver this talk and for their very generous hospitality.

\end{document}